\documentclass{article}
\usepackage{graphicx}

\newcommand\rs[1]{_\mathrm{#1}}   
\newcommand\U[1]{{\,\rm #1}}
\newcommand\E[1]{\times10^{#1}}
\newcommand{\ARAA}{Ann. Rev. Astr. Aph.}
\newcommand{\AandA}{Astr. \& Aph.}
\newcommand{\ApJ}{Ap. J.}
\newcommand{\ApJS}{Aph. J. Suppl.}
\newcommand{\mnras}{Mon. Not. Roy. Ast. Soc.}
\newcommand{\sovastl}{Sov. Astr. Let.}
\newcommand{\revmp}{Rev. Mod. Phys.}
\newcommand{\kfnt}{Kinem. and Phys. of Celestial Bodies}
\newcommand\timesss[2]{\makebox[1.1cm][l]{#1}\parbox[t]{0.93\textwidth}{#2}\\}

\begin{document}

\begin{centering}
\noindent{\LARGE\bf On the Transition of the Adiabatic Supernova Remnant to the Radiative Stage
in a Nonuniform Interstellar Medium}\\
Oleh Petruk\\
Institute for Applied Problems in Mechanics and Mathematics\\
3-b Naukova St., 79053 Lviv, Ukraine\\
petruk@astro.franko.lviv.ua\\
\end{centering}

\vspace{0.5cm}

{\bf Abstract.} {\small Methods for estimation of different reference times which  appear
in the description of transition of a strong adiabatic shock into the radiative era are
reviewed. The need for consideration of an additional transition subphase in between the
end of the adiabatic era and the beginning of the radiative ``pressure-driven snowplow''
stage for a shock running in the uniform or nonuniform medium is emphasized. This could
be of importance in particular for studying of the interaction of supernova remnants
(SNRs) with molecular clouds and therefore for understanding the processes of the cosmic
ray production in such systems. The duration of this subphase -- about 70\% of SNR age at
its beginning -- is almost independent of the density gradient for media with increasing
density and is longer for higher supernova explosion energy and for smaller density in
the place of explosion. It is shown as well that if the density of the ambient medium
decreases then the cooling processes could differ from the commonly accepted scenario of
the ``thin dense radiative shell'' formation. This property should be studied in the
future because it is important for models of nonspherical SNRs which could be only
partially radiative. }



\section{Introduction}

Physical processes accompanying the evolution of supernova  remnants (SNRs) is a complex
system. It is almost impossible to account for all of them in a single model of SNR.
Therefore, the whole evolution of SNR from a supernova explosion until the mixing of a
very old object with the interstellar matter is divided on a number of the model phases
(e.g. \cite{Shklovsky-68,Woltjer-72, Lozinskaya-92}): the free-expansion, adiabatic,
radiative and dissipation stages. There are some physical processes important during a
given stage, some others could be neglected. Such an approach allows for rather simple
analytical description of SNR evolution during each phase.

The role of radiative losses, which is negligible in the  adiabatic phase of SNR
evolution, becomes more and more promiment with time. They are so important in old SNRs,
that they essentially modify the dynamics of such SNRs. Theoretical systematization of
timescales and the role of different physical process in cooling of adiabatic SNR was
first reviewed in \cite{Cox-72}. In general, the transition to the radiative stage can be
studied numerically, by following the history of the shocked flow as it is done e.g. by
\cite{Chev-74,Cioffi-et-al-88,Falle-81,Mansf-Salp-74,Franko-et-al-94,Blondin}. The
analytical treatments are of great importance as well, e.g.
\cite{Oort-51,McKee-Ostr-77,Blinn-et-al-82,Silich-Pasko-86,Ostr-McKee-88,Band-Petr-04}.

The physical processes in the radiative blast wave, namely,  quick cooling of an incoming
flow and formation of the thin dense cold shell which moves due to the pressure of
internal gas makes the so called ``pressure-driven snowplow'' (PDS) model within the
``thin-layer'' approximation to be  adequate for description of this stage of SNR
evolution \cite{Zel-Raizer-66,Ostr-McKee-88,BK-Syl}.

The PDS model was introduced by McKee \& Ostriker; their  analytical solution
\cite{McKee-Ostr-77,Ostr-McKee-88} widely used for the description of evolution of the
radiative shell gives a power-law dependence $R\propto t^{m}$ (where $t$ is age and $R$
is a position of the shock) with constant $m$ (which equals to $2/7$ for the uniform
medium). However, numerical studies cited above give something different values of the
deceleration parameter $m$ (defined as $m=d\ln R/d\ln t$), namely $\approx 0.33$
\cite{Chev-74,Blondin}. We have shown analytically in \cite{Band-Petr-04} that the
evolution of the radiative shell is given by variable $m$ and that the discrepancy
between the analytical and numerical results is only apparent. In fact, the usage of
McKee \& Ostriker analytical solution assumes that SNR has already reached the asymptotic
power-law regime with constant value of $m=2/7$. The time needed to reach this asymptotic
regime is however long compared to the SNR age.

It is common for an approximate theoretical description  of SNR evolution to simply
switch from the adiabatic solution to the PDS radiative one at some moment of time.
However, we stress in this paper the result visible also in previous calculations,
namely, the need for an intermediate transition subphase between the adiabatic and
radiative stages, with duration more than a half of SNR age it has at the time when
radiative losses of gas passing through the shock begins to be prominent. Thus the
radiative era which begins after the end of the adiabatic one, have to be divided on two
phases: the transition subphase, when the radiative losses become to modify dynamics and
to lead to the formation of the thin radiative shell, and the PDS stage when one can
apply the PDS analytic solution. In the present paper the role of nonuniform interstellar
medium on the duration of the transition subphase is considered.

\section{Transition to the radiative phase}\label{sect1}

\subsection{Definitions of different reference times}

Let us consider the spherical shock motion in the medium  with the power-law density
variation $\rho\rs{o}(R)=AR^{-\omega}$, where $A$ and $\omega$ are constant; indexes
``o'' and ``s'' refer hereafter to the pre- and post-shock values. The dynamics of the
adiabatic shock in such a medium is given by Sedov solutions \cite{Sedov-59} where the
shock velocity $D\propto R^{-(3-\omega)/2}$ and $R\propto t^{2/(5-\omega)}$.

Moving through medium, the shock decelerates if the ambient  density distribution
increases or does not quickly decrease ($\omega<3$). The shock temperature
$T\rs{s}\propto D^2$ decreases with time as well. Starting from some age $t\rs{low}$ when
$T\rs{s}=T\rs{low}\sim 3\E{7}\U{K}$, which corresponds to the minimum of the cooling
function $\Lambda(T)$, the radiative losses of shocked plasma are more and more prominent
with falling of $T$ (Fig.~\ref{fig-a}). The maximum in the energy losses is when the
shock temperature $T\rs{s}=T\rs{hi}\sim 2\E{5}\U{K}$, the corresponding Sedov time (i.e.
calculated under the assumption that the shock is adiabatic up to this time) is
$t\rs{hi}$.

There is a number of reference times in between $t\rs{low}$  and $t\rs{hi}$
\cite{Cox-72,Cox-Anderson-82,Blondin}. Once a parcel of gas is shocked its temperature
changes due to expansion and cooling $\dot{T}\rs{a}=\dot{T}\rs{a,exp}+\dot{T}\rs{a,rad}$,
where the dot marks the time derivative. One may define the ``dynamics-affected'' time
$t\rs{dyn}$ by the equation
\begin{equation}
 \dot{T}\rs{a,exp}(t\rs{dyn})=\dot{T}\rs{a,rad}(t\rs{dyn}).
 \label{eq-tdyn}
\end{equation}
If a fluid element is shocked after this time, its temperature  decreases faster due to
radiation than as a consequence of expansion. At other time $t\rs{sag}$, the radiative
cooling begins to affect the temperature distribution inside the shock. When the rate of
change of the shock temperature $\dot{T}\rs{s}$ begin to be less than $\dot{T}\rs{a}$,
the temperature downstream of the shock will sag rather than rise. Thus the equation for
$t\rs{sag}$ is
\begin{equation}
 \dot{T}\rs{s}(t\rs{sag})=\dot{T}\rs{a}(t\rs{sag}).
 \label{eq-tsag}
\end{equation}

Radiative losses cause the faster -- comparing to the adiabatic  phase -- deceleration of
the forward shock. This faster deceleration begins to be prominent around the
``transition age'' $t\rs{tr}$ when the shock pressure decrease due to the radiative
losses becomes to be effective. Then, the shocked gas radiates away its energy rather
quickly, cools till the temperature $T\sim 10^4\ {\rm K}$ and forms a dense shell. The
formation of the shell is completed around the ``time of shell formation'' $t\rs{sf}$
which is larger than $t\rs{tr}$; the latter which marks the end of adiabatic era. After
$t\rs{sf}$ the thermal energy of all swept-up gas is rapidly radiated and the thin dense
shell expansion is caused by the thermal pressure of the interior.

The time $t\rs{low}$ is given by the equation
\begin{equation}
 T\rs{s}(t\rs{low})=T\rs{low}.
 \label{eq-tlow}
\end{equation}
A similar equation defines the time $t\rs{hi}$
\begin{equation}
 T\rs{s}(t\rs{hi})=T\rs{hi},
 \label{eq-thi}
\end{equation}
which was suggested to be a measure of $t\rs{tr}$ \cite{Blinn-et-al-82,Silich-Pasko-86}.
However, as we shall demonstrate later, the post-shock  temperature of plasma at
$t\rs{tr}$ is of order $10^6\U{K}>T\rs{hi}$ and $t\rs{hi}$ is larger than $t\rs{tr}$ in
about $3.5$ times (Sect.~\ref{sect-times-uni}). Therefore it is not correct to calculate
the ``highest-losses'' SNR age with the shock motion law valid during the adiabatic era.

A simple approach to locate $t\rs{tr}$ bases on the comparison  of the radiative losses
with the initial thermal energy of the shocked fluid \cite{Blondin}. A shocked fluid
element cools during the cooling time $\Delta
t\rs{cool}\propto\epsilon(T\rs{s},\rho\rs{s})/\Lambda(T\rs{s},\rho\rs{s})$, where
$\epsilon=(\gamma-1)^{-1}\rho\rs{s}k\rs{B}T\rs{s}/\mu m\rs{p}$ is its initial thermal
energy density, $\gamma$ is the adiabatic index, $k\rs{B}$ is the Boltzman constant,
$m\rs{p}$ is the proton mass. During the adiabatic phase the cooling time is larger than
SNR age $t$. The radiative losses may be expected to modify dynamics when the cooling
time $\Delta t\rs{cool}\leq t$. In such approach the transition time is a solution of
equation
\begin{equation}
 t\rs{tr}=\Delta t\rs{cool}(t\rs{tr}).
 \label{eq-t-tr}
\end{equation}
Let us assume that the cooling function $\Lambda\propto n^2T^{-\beta}$ with  $\beta>0$
and $n$ is the hydrogen number density, then $\Delta t\rs{cool}\propto
n\rs{o}^{-1}T\rs{s}^{1+\beta}\propto t^{-6(1+\beta)/5}$ with the use of Sedov solutions
for uniform medium. For the shock running in the power-law density distribution, the
upstream hydrogen number density and the post-shock temperature at time $t$ is
\begin{equation}
 n\rs{o}\propto t^{-2\omega/(5-\omega)}, \qquad T\rs{s}\propto t^{-2(3-\omega)/(5-\omega)}.
 \label{n-t-T-t}
\end{equation}
Therefore
$\Delta t\rs{cool}\propto t^{-\eta}$ with $\eta=\big(2(3-\omega)(1+\beta)-2\omega)/(5-\omega)\big)$ for such density distribution. For $\beta=1/2$ the index $\eta$ is the same as found in \cite{Franko-et-al-94}.

The way to estimate the time of the shell formation $t\rs{sf}$ was  suggested in
\cite{Cox-Anderson-82,Cox-86}. If an element of gas was shocked at time $t\rs{s}$ then
the age of SNR will be $t\rs{c}=t\rs{s}+\Delta t\rs{cool}(t\rs{s})$ when it cools down.
The minimum of the function $t\rs{c}(t\rs{s})$ has the meaning of SNR age when the first
element of gas cools and is called ``SNR cooling time'' $t\rs{cool}$. Let $t_1$ be the
time when the shock encountered the fluid element which cools first. If so,
$t\rs{c}=t_1(t\rs{s}/t_1)+\Delta t\rs{cool}(t_1)(t\rs{s}/t_1)^{-\eta}$. Setting
$dt\rs{c}/dt\rs{s}|_{t\rs{s}=t_1}=0$ one obtain that
\begin{equation}
 t\rs{cool}=(1+\eta)\Delta t\rs{cool}(t_1),
 \label{eq-t-sf}
\end{equation}
\begin{equation}
 \frac{t\rs{cool}}{t_1}=\frac{1+\eta}{\eta},
 \label{ratio-tcool-t1}
\end{equation}
The cooling time $t\rs{cool}>t_1$ by the definition, therefore it  must be that $\eta>0$.
This is the case for
\begin{equation}
 \omega<3(1+\beta)/(2+\beta);
 \label{cond-beta}
\end{equation}
that is $\omega<2\ (9/5)$ for $\beta=1\ (1/2)$.
The equation
\begin{equation}
 t_1=\eta\Delta t\rs{cool}(t_1)
 \label{eq-t1}
\end{equation}
is more suitable for practical use than (\ref{eq-t-sf}). If the medium is uniform then
$t\rs{cool}=17t_1/12$ for $\beta=1$  and $t\rs{cool}=14t_1/9$ for $\beta=1/2$.

The ``SNR cooling time'' $t\rs{cool}=\min(t\rs{c})$ was initially  suggested to be taken
as the time of the shell formation. Numerical experiments for shock in the uniform medium
suggest that $t\rs{sf}$ is a bit higher (of order 10\%) than $t\rs{cool}$
\cite{Cox-et-al-99} and the reason of this could be that the compression of the shell is
also effective after cooling of the first element  and takes additional time.

Another point is that the solution for adiabatic shock used in (\ref{n-t-T-t}) might not
formally be applicable there because $t_1>t\rs{tr}$ (see Eq.~(\ref{t1_ttr})). We believe
however that the level of accuracy in estimation of $t\rs{tr}$, the small difference
between $t\rs{tr}$ and $t_1$ (about $30\%$ in the case of uniform medium,
Sect.~\ref{sect-times-uni}) as well as close values of $t\rs{cool}$ and $t\rs{sf}$ allow
one to use the Sedov solution in (\ref{n-t-T-t}) and to assume $t\rs{sf}\approx
t\rs{cool}$.

We would like to note once more that the transition time $t\rs{tr}$  is an approximate
estimation on the end of the adiabatic stage and beginning of the radiative era, while
the time of the shell formation $t\rs{sf}$ marks the time when one can start to use the
PDS model where hot gas pushes the cold dense shell\footnote{The PDS analytical solutions
which describe the evolution of SNR after the shell formation time are presented in
\cite{Blinn-et-al-82,Silich-Pasko-86,Band-Petr-04} for uniform ISM and in
\cite{Silich-Pasko-86} for ISM with power-law density variation.}. The structure of the
flow re-structurises and the shell forms during the transition subphase given  by the
time interval $(t\rs{tr},t\rs{sf})$. We shall demonstrate later that the ratio
$t\rs{sf}/t\rs{tr}$ with $t\rs{tr}$ given by (\ref{eq-t-tr}) and $t\rs{sf}$ by
(\ref{ratio-tcool-t1}) is always larger than unity (see Eq.~(\ref{tsf_ttr})) and that the
transition subphase is not short as it is generally assumed.

One more time, namely the ``intersection time'' $t\rs{i}\in (t\rs{tr},t\rs{sf})$ was
introduced in \cite{Band-Petr-04}, as a time when two functions -- the adiabatic
dependence $R=R(t)$ (valid before $t\rs{tr}$) and the PDS dependence
$R\rs{sh}=R\rs{sh}(t)$ (valid after $t\rs{sf}$) -- intersect being extrapolated into the
transition subphase. This intersection time could be useful in some tasks when the  level
of accuracy is such that one may sharply switch from the adiabatic solution to the
radiative one without consideration of the transition subphase.

\subsection{Cooling time}

The expression
\begin{equation}
 \Delta t\rs{cool}={\epsilon(T\rs{s},\rho\rs{s})\over\Lambda(T\rs{s},\rho\rs{s})}
 \label{delta-cool}
\end{equation}
used in \cite{Blondin} to calculate the cooling time, equates the energy losses $\Lambda
\Delta t\rs{cool}$ with initial  thermal energy density $\epsilon\rs{s}$ of a fluid
element under condition that the density and temperature of this element are constant.
More detailed model should account for the density and temperature history during $\Delta
t\rs{cool}$. Namely the above equation should be replaced with a differential one:
\begin{equation}
 d\epsilon/dt=-\Lambda(T,\rho).
 \label{dif-L}
\end{equation}
The total internal energy $U=\epsilon V$ of gas within the volume  $V$ changes as
$dU=TdS-PdV$ where $S$ is entropy and $P$ is pressure. The evolution of the thermal
energy per unit mass $E=\epsilon/\rho$ is therefore
\begin{equation}
 {\partial E\over \partial t}-{P\over \rho^2}\left({\partial \rho\over \partial t}\right)=
 T{\partial s\over \partial t}
 \label{eq6}
\end{equation}
where $s=(3k\rs{B}/2m\rs{p}\mu)\ln\left(P/\rho^{\gamma}\right)$ is  the entropy per unit
mass ($m\rs{p}$ is the mass of proton, $\mu$ is the mean particle weight). So,
Eq.~(\ref{dif-L}) becomes
\begin{equation}
 T{\partial s\over \partial t}=-{\Lambda(T,\rho)\over\rho},
 \label{dif-L-Kahn}
\end{equation}
here the temperature $T$, density $\rho$, pressure $P$, energy $E$  are functions of
Lagrangian coordinate $a$ and time $t$.

As it follows from (\ref{dif-L-Kahn}) and the definition of $s$, the  time $\Delta
t\rs{cool}$ may be also defined as a time taken for the adiabat $P/\rho^{\gamma}$ to fall
to zero. Kahn \cite{Kahn-1976} have found an interesting result. Namely, if
\begin{equation}
 \beta=\frac{2-\gamma}{\gamma-1}
 \label{beta-uniq}
\end{equation}
(that is $\beta=1/2$ for $\gamma=5/3$)
then one can derive $\Delta t\rs{cool}$ from (\ref{dif-L-Kahn})
independently of the density and temperature history:
\begin{equation}
 \Delta t\rs{cool}^{\rm Kahn}={\epsilon(T\rs{s},\rho\rs{s})\over (\beta+1)\Lambda(T\rs{s},\rho\rs{s})}.
 \label{cox-t-cool}
\end{equation}
It can be checked that the same solution may be obtained from
(\ref{eq6})-(\ref{dif-L-Kahn}) for any $\beta$ if one assume that the gas is not doing
work during $\Delta t\rs{cool}$ that is equivalent to putting $\partial\rho/\partial t=0$
in (\ref{eq6}). However, the density of fluid is not expected to be constant. In such
situation one should solve the full set of the hydrodynamic equations which can be
performed only numerically, while we are interested in a rather simple analytical
estimation on cooling time for a general $\beta$. Therefore it is more suitable to use
the estimation (\ref{delta-cool}) for the cooling time which follows just from comparison
of the radiative losses with the initial energy. We shall see later that such approach
describe the shock dynamics rather well (Fig.~\ref{fig-m}).

\subsection{Equations for the reference times}

Let us write equations for $t\rs{tr}$ and $t\rs{sf}$ for shock  in nonuniform medium. We
assume hereafter $\beta=1$. Note that all the rest formulae can easily be modified if one
uses $\beta$ which coincides with a value given by  (\ref{beta-uniq}); namely, as it
follows from comparison of (\ref{cox-t-cool}) and (\ref{delta-cool}), ${\cal T}$ in
(\ref{deltat}) have to be simply divided by $\beta+1$.

If the cooling function for a fluid is approximately $\Lambda=CT^{-\beta}n_en_H$, where
$C$ is a constant, then (\ref{delta-cool}) yields
\begin{equation}
 \Delta t\rs{cool}={\cal T}{T\rs{s}^{1+\beta}\over n\rs{o}(R)}
 \quad \mathrm{where}\quad {\cal T}={k\rs{B}\mu_e\over C\mu(\gamma+1)},
 \label{deltat}
\end{equation}
$\mu_e$ is the mean mass of particle per one electron in terms  of the proton mass (i.e.
$\rho=\mu_en_em\rs{p}=\mu n m\rs{p}$). The transition time $t\rs{tr}$ is a solution of
equation (\ref{eq-t-tr}):
\begin{equation}
 t\rs{tr}={\cal T} {T_{s}(t\rs{tr})^{1+\beta}\over n\rs{o}(R(t\rs{tr}))},
 \label{eq11}
\end{equation}
where the dependencies $T\rs{s}(t)$, $R(t)$ are those valid on the adiabatic phase. The
time $t_1$ can be estimated from (\ref{eq-t1}):
\begin{equation}
 t_1=\eta{\cal T}{T_{s}\left(t_1\right)^{1+\beta}\over
  n\rs{o}\left(R(t_1)\right)}.
 \label{eq2-t1}
\end{equation}
Now the SNR cooling time $t\rs{cool}$ and the time of the shell  formation
$t\rs{sf}\approx t\rs{cool}$ is given by (\ref{ratio-tcool-t1}).
The estimations for the transition and the shell formation times  are somewhat different
in the literature because of different  ways used to find the cooling time $\Delta
t\rs{cool}$ and to approximate the cooling function $\Lambda(T)$.

For the adiabatic shock the rate of change of the shock temperature is
\begin{equation}
 \dot{T}\rs{s}=-{2(3-\omega)\over 5-\omega}\frac{T\rs{s}}{t}.
\end{equation}
Close to the shock, the fluid temperature in Sedov solution  \cite{Sedov-59} is
approximately
\begin{equation}
 \frac{T(a)}{T\rs{s}}\approx \left(\frac{a}{R}\right)^{-\kappa(\gamma,\omega)},
\end{equation}
where $a$ is Lagrangian coordinate. The value of $\kappa$ is given by
\begin{equation}
 \kappa= \left(-{a\over T(a)}{\partial T(a)\over \partial a}\right)_{a=R}.
\end{equation}
where $T(a)$ is the profile from Sedov solutions. It is $\kappa=1-3\omega/4$ for
$\gamma=5/3$ (see Appendix). Now we may find that the temperature in a given fluid
element  $a$ changes due to expansion as
\begin{equation}
 \dot{T}\rs{a,exp}\approx -{2(3-\omega-\kappa)\over 5-\omega}\frac{T(a)}{t}.
\end{equation}
The rate $\dot{T}\rs{a,rad}$ due to cooling follows from $dE/dt=-\Lambda/\rho$:
\begin{equation}
 \dot{T}\rs{a,rad}=-\frac{\gamma-1}{\gamma+1}{\cal T}^{-1}n\rs{H}(a)T(a)^{-\beta}.
\end{equation}
Now we have to compare the above rates at the time $t\rs{s}$,  i.e. at the time when the
parcel of fluid was shocked. The coordinate $a=R(t\rs{s})$ by the definition. Thus
Eq.~(\ref{eq-tdyn}) rewrites:
\begin{equation}
 t\rs{dyn}={2(3-\omega-\kappa)\over 5-\omega}\Delta t\rs{cool}(t\rs{dyn}).
 \label{eq2-tdyn}
\end{equation}
Similarly, the equation for $t\rs{sag}$ follows from (\ref{eq-tsag}):
\begin{equation}
 t\rs{sag}={2\kappa\over 5-\omega}\Delta t\rs{cool}(t\rs{sag}).
 \label{eq2-tsag}
\end{equation}

As one can see, the most of reference times are given by the  equations of the form
\begin{equation}
 t\rs{*}=K\Delta t\rs{cool}(t\rs{*}),
\end{equation}
where $t\rs{*}$ is a given reference time and $K$ is  corresponding constant. It may be
shown that the solution of such equation may be found as
\begin{equation}
 t\rs{*}=K^{1/(1+\eta)}t\rs{tr}.
 \label{t*-ttr}
\end{equation}
The Sedov radius of the shock at this time is
$R\rs{*}=K^{2/\left((5-\omega)(1+\eta)\right)}R\rs{tr}$.

\begin{figure}
\centering
\includegraphics{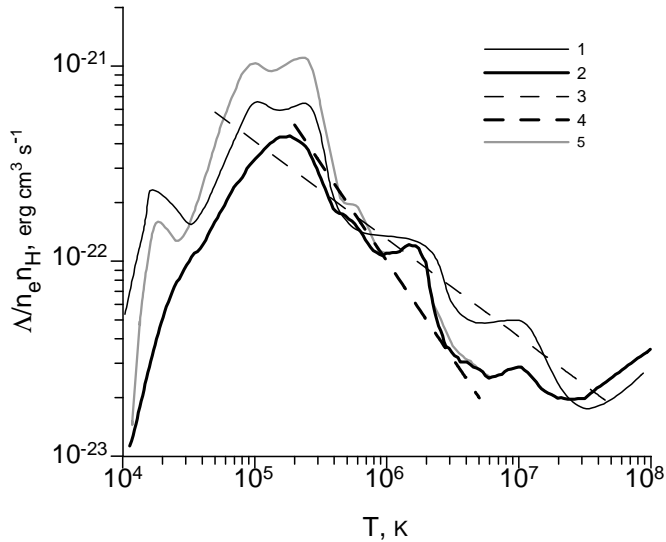}
\caption{Equilibrium (line 1) \cite{Raymond-Cox-Smith-76}
and nonequilibrium (line 2) \cite{Suth-Dopita-93} cooling functions, used in the literature
to study the transition of SNRs into the radiative phase, and
approximations (\ref{L-kahn-app}) (line 3) and
(\ref{L-blond-app}) (line 4). The equilibrium cooling function from \cite{Suth-Dopita-93} is
also shown for comparison (line 5).
}
\label{fig-a}
\end{figure}

\subsection{The cooling function}

There are two choices of $\beta$ in the literature, namely $1$ and $1/2$. The first case
is used for {\em nonequilibrium} cooling  model \cite{Suth-Dopita-93} where the cooling
function for  plasma with solar abundance may be approximated as \cite{Blondin}
\begin{equation}
 \Lambda=10^{-16}n_en_HT^{-1}\ {\rm erg\ cm^{-3}\ s^{-1}}.
 \label{L-blond-app}
\end{equation}
This approximation is valid for range of temperatures  $T=(0.2-5)\E{6}\U{K}$ which is
important for description of transition into the radiative phase. Another possibility is
to use the {\em equilibrium} cooling model as it was done in
\cite{Kahn-1976,Cox-Anderson-82,Franko-et-al-94,Maciejewski-Cox-99,Cox-et-al-99}. In this
case the approximate proportionality $\Lambda\propto T^{-1/2}$ is a reasonable one, e.g.
for results on the cooling of the collisional equilibrium plasma from
\cite{Raymond-Cox-Smith-76,Smith-et-al-1996}; the actual approximation
\begin{equation}
 \Lambda=1.3\E{-19}n_en_HT^{-1/2}\ {\rm erg\ cm^{-3}\ s^{-1}}
 \label{L-kahn-app}
\end{equation}
is written for plasma with almost the same abundance as above and is valid for
$T=(0.05-50)\E{6}\U{K}$ \cite{Kahn-1976}.

Different cooling functions are compared with their approximations on Fig.~\ref{fig-a}.
At lower temperatures, the nonequilibrium cooling is less effective in energy losses than
the equilibrium one (compare lines 2 and 5). This is because the cooling rate for
temperatures higher than $\sim 3\E{7}\U{K}$ is mostly due to free-free emission while
below this temperature the  cooling is mostly due to the line emission from heavy
elements (most heavy elements are completely ionized above $\sim 3\E{7}\U{K}$). Under
nonequilibrium ionization conditions the ions are underionized because electrons are much
colder than ions and thus there is less emission from ions
\cite{Franko-et-al-94,Hamilton-Sarazin-Chevalier} (see also Fig.~18 in
\cite{Suth-Dopita-93}).

\section{Reference times and transition subphase}

\subsection{Shock in a uniform ISM}\label{sect-times-uni}

Let us compare the sequence of different reference times  with numerical calculations
\cite{Blondin} of transition of the adiabatic shock into the radiative era, on example of
the shock motion in the uniform ambient medium. Let us consider the same parameters as in
\cite{Blondin}, namely $\gamma=5/3$, $\beta=1$, the same abundance ($\mu=0.619$,
$\mu\rs{e}=1.18$, $\mu\rs{H}=1.43$) as well as assume $t\rs{sf}=t\rs{cool}$ and use
(\ref{delta-cool}) for calculation of $\Delta t\rs{cool}$.

If shock wave moves in the uniform medium, then -- with  the use of Eq.~(\ref{eq11}) --
the transition time is
\begin{equation}
  t\rs{tr}=2.84\E{4}E_{51}^{4/17}n\rs{o}^{-9/17}\U{yr}
    \label{ttran}
\end{equation}
where $E_{51}=E\rs{SN}/(10^{51}\U{erg})$. The gas element which first cools (at
$t\rs{cool}$) was  shocked at $t_1$ which follows from Eq.~(\ref{eq2-t1}):
\begin{equation}
  t\rs{1}=3.67\E{4}E_{51}^{4/17}n\rs{o}^{-9/17}\U{yr}.
    \label{t1}
\end{equation}
The time of the shell formation is given by Eq.~(\ref{ratio-tcool-t1}):
\begin{equation}
  t\rs{sf}=5.20\E{4}E_{51}^{4/17}n\rs{o}^{-9/17}\U{yr},
    \label{tsf}
\end{equation}
so that $t\rs{sf}/t\rs{tr}=1.83$. The time when the radiative losses of the shocked gas
reach their minimum is (\ref{eq-tlow}):
\begin{equation}
 t\rs{low}=1.60\E{3}T\rs{3e7}^{-5/6}E_{51}^{1/3}n\rs{o}^{-1/3}\U{yr}
 \label{tloss}
\end{equation}
where $T\rs{3e7}=T\rs{low}/(3\E{7}\U{K})$. Under assumption that  radiative losses does
not change the shock dynamics till $t\rs{hi}$, with the use of Sedov solutions for the
shock motion one have from Eq.~(\ref{eq-thi}) that
\begin{equation}
 t\rs{hi}=1.04\E{5}T\rs{2e5}^{-5/6}E_{51}^{1/3}n\rs{o}^{-1/3}\U{yr}
 \label{thi}
\end{equation}
where $T\rs{2e5}=T\rs{hi}/(2\E{5}\U{K})$.
The fluid temperature drops faster due to cooling than due to expansion from time
\begin{equation}
  t\rs{dyn}=2.66\E{4}E_{51}^{4/17}n\rs{o}^{-9/17}\U{yr}.
    \label{tdyn}
\end{equation}
The time when one may expect to have the temperature decrease  downstream close to the
shock is
\begin{equation}
  t\rs{sag}=2.17\E{4}E_{51}^{4/17}n\rs{o}^{-9/17}\U{yr}.
    \label{tsag}
\end{equation}

The Sedov solutions give at time $t\rs{tr}$ the shock radius
$R\rs{tr}=19E_{51}^{5/17}n\rs{o}^{-7/17}\U{pc}$, the  shock velocity $D\rs{tr}=260\
E_{51}^{1/17}n\rs{o}^{2/17}\U{km/s}$, the post-shock temperature $T\rs{tr}=0.95\cdot
10^6\ E_{51}^{2/17}n\rs{o}^{4/17}\U{K}$ and the swept up mass $M\rs{tot}(t\rs{tr})=10^3\
E_{51}^{15/17}n\rs{o}^{-4/17}\U{M_\odot}$.

\begin{figure}
\centering
\includegraphics{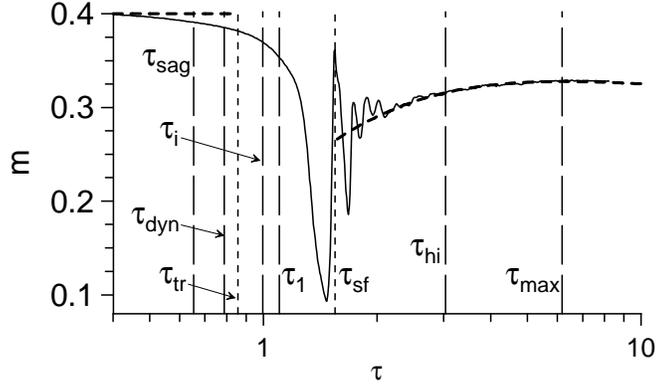}
\caption{The evolution of the deceleration parameter $m$ and different reference times
for the shock motion in the uniform medium. Solid line -- numerical calculations
\cite{Blondin}, thick dashed lines -- Sedov solution (till $\tau\rs{tr}$) and analytical
solution \cite{Band-Petr-04} (after $\tau\rs{sf}$). The dimensionless reference times are
$\tau\rs{sag}=0.654$, $\tau\rs{dyn}=0.802$, $\tau\rs{tr}=0.855$,  $\tau\rs{i}=1.01$,
$\tau_1=1.10$, $\tau\rs{sf}=1.57$, $\tau\rs{low}=0.047$, $\tau\rs{hi}=3.03$. The function
$m(\tau)$ reaches his maximum at radiative phase at $\tau\rs{max}=6.18$
\cite{Band-Petr-04}. } \label{fig-m}
\end{figure}

The above reference times are shown on Fig.~\ref{fig-m} together  with evolution of the
deceleration parameter $m(\tau)$ calculated numerically \cite{Blondin}. The analytical
solutions for the adiabatic \cite{Sedov-59} and the  radiative shock \cite{Band-Petr-04}
are also shown. Numerical result is found for supernova energy $E\rs{SN}=10^{51}\U{erg}$
and interstellar hydrogen number density $n\rs{o}=0.84\U{cm^{-3}}$. With these values,
the times are $t\rs{sag}=2.4\E{4}\U{yr}$, $t\rs{dyn}=2.9\E{4}\U{yr}$,
$t\rs{tr}=3.1\E{4}\U{yr}$,  $t_1=4.0\E{4}\U{yr}$, $t\rs{sf}=5.7\E{4}\U{yr}$,
$t\rs{low}=1.7\E{3}\U{yr}$, $t\rs{hi}=1.1\E{5}\U{yr}$; the intersection time is
$t\rs{i}=3.6\E{4}\U{yr}$  \cite{Band-Petr-04}. The function $m(\tau)$ reaches his maximum
during the radiative stage at $t\rs{max}=2.3\E{5}\U{yr}$ \cite{Band-Petr-04}. Results on
Fig.~\ref{fig-m} are presented in terms of the dimensionless time $\tau=t/\tilde t$
because the analytical solutions allow for scaling (numerical results for various input
parameters differs by oscillation transient only; see e.g. Fig.~8 in \cite{Blondin}). The
dimensional scale for time determined from fitting of analytical and numerical results is
$\tilde t=3.6\E{4}\U{yr}$ \cite{Band-Petr-04}.

It is apparent from Fig.~\ref{fig-m} that the transition time  $t\rs{tr}$ is a reasonable
estimation for the end of the  adiabatic stage while $t\rs{sf}$ could be the time when
one can start to use the radiative solutions \cite{Band-Petr-04} coming from the PDS
model of McKee \& Ostriker \cite{McKee-Ostr-77}. The duration of the intermediate
transition subphase is $(\tau\rs{sf}-\tau\rs{tr})/\tau\rs{tr}=0.83$ times the age of SNR
at the end of the  adiabatic stage, i.e. almost the same as duration of the adiabatic
stage itself. This means that there is a strong need for a theoretical model which
describe evolution of SNR in this subphase.

\begin{figure}
\centering
\includegraphics{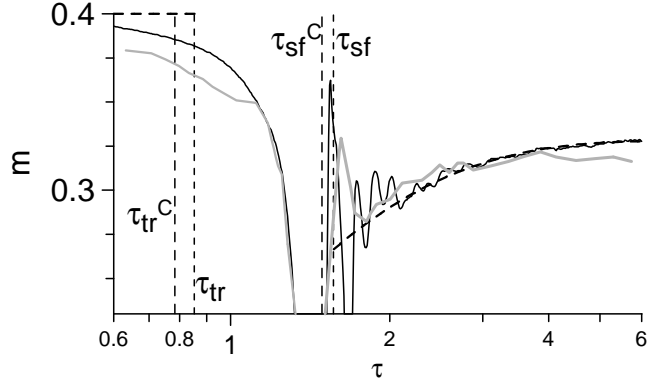}
\caption{Numerical calculation of evolution of the  deceleration parameter $m$ from
\cite{Blondin} (thin black line) and from \cite{Cioffi-et-al-88} (thick gray line). The
transition and shell formation times from \cite{Cioffi-et-al-88} are marked by ``C''. }
\label{fig-mCioffi}
\end{figure}

For estimation of reference times, a number of authors
\cite{Cox-Anderson-82,Cox-86,Cioffi-et-al-88,Franko-et-al-94,Cox-et-al-99} keep a bit
different approach from that used above, namely they use the approximation of the
equilibrium cooling function with $\beta=1/2$ and the Kahn solution for cooling time
(\ref{cox-t-cool}). Let us compare the results of this approach with those obtained
above. The evolution of the deceleration parameter in the refereed approach is presented
in \cite{Cioffi-et-al-88}. There is also the same definition of the time of the shell
formation $t\rs{sf}=t\rs{cool}$. The estimation is
$t\rs{sf,C}=4.31\E{4}E_{51}^{3/14}n\rs{o}^{-4/7}\U{yr}$ for their abundance and the
cooling function (\ref{L-kahn-app}). For the parameters used in the numerical
calculations $E\rs{51}=0.931$ and $n\rs{o}=0.1\U{cm^{-3}}$ the time is
$t\rs{sf,C}=1.58\E{5}\U{yr}$ while with the use of our Eq.~(\ref{tsf}) we obtain
$t\rs{sf}=1.73\E{5}\U{yr}$. The both estimations are close. Analytical solutions shows
that, before $t\rs{tr}$ and after $t\rs{sf}$, the evolution of dynamic parameters of the
shock can be expressed in a dimensionless form, i.e. independently of $E_{51}$ and
$n\rs{o}$.  The behavior of the shock velocity depends however on these parameters during
the transition subphase; the difference is in the frequency of oscillations (Fig.~8 in
\cite{Blondin}). Nevertheless, as one can see from this figure, the strong deceleration
of the shock right after $t\rs{tr}$ up to the first minimum is almost the same for
different parameters, i.e. can also be scaled. We use this property in order to find the
scale factor $\tilde t$ for calculations being done in \cite{Cioffi-et-al-88}. Namely,
the fit of curve $m(\tau)$ from \cite{Cioffi-et-al-88} to that of \cite{Blondin} (within
the time interval from $t\rs{tr}$ to the first minimum) gives $\tilde
t\rs{C}=1.05\E{5}\U{yr}$. The both calculations of the transition to the radiative stage
agree rather well as it may be seen on Fig.~\ref{fig-mCioffi}. The dimensionless times
for results in \cite{Cioffi-et-al-88} are: the shell formation time
$\tau\rs{sf,C}=t\rs{sf,C}/\tilde t_\mathrm{C}=1.51$ and the transition time (as it is
follows from (\ref{tsf_ttr}))  $\tau\rs{tr,C}=\tau\rs{sf,C}/1.92=0.785$.
Fig.~\ref{fig-mCioffi} shows that the both approaches for localization of the limits of
the transition subphase -- with the use of the nonequilibrium-ionization cooling function
(\ref{L-blond-app}) and the simple estimation for $\Delta t\rs{cool}$ (\ref{delta-cool})
\cite{Blondin} or with the equilibrium cooling function (\ref{L-kahn-app}) together with
Kahn solution for $\Delta t\rs{cool}$ (\ref{cox-t-cool}) \cite{Cioffi-et-al-88} -- give
almost the same estimations.

\subsection{Shock in a medium with a power-law density variation}

\begin{figure}
\centering
\includegraphics{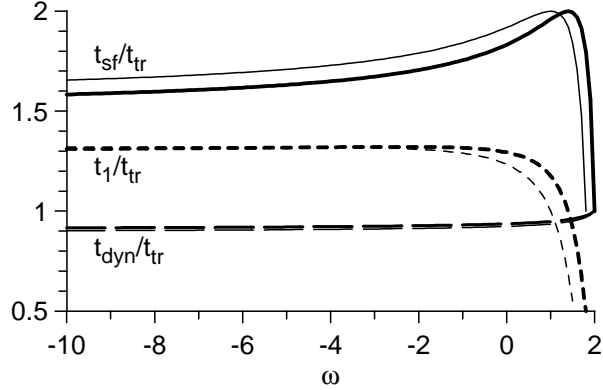}
\caption{The ratios of times for $\beta=1$ (thick lines)  and $\beta=1/2$ (thin lines) as
it is obtained from (\ref{tsf_ttr}) and (\ref{t1_ttr}). } \label{fig-b}
\end{figure}

Let us now consider the shock motion in the ambient medium  with the power-law density
variation $\rho^o(R)=AR^{-\omega}$. With the use of (\ref{eq11}), (\ref{eq2-t1}),
(\ref{ratio-tcool-t1}), (\ref{n-t-T-t}) and the definition $t\rs{sf}=t\rs{cool}$ one can
show that the duration of the transition subphase is given by
\begin{equation}
 {t\rs{sf}\over t\rs{tr}}={t\rs{cool}\over t\rs{tr}}={1+\eta\over\eta^{\eta/(1+\eta)}}\ .
 \label{tsf_ttr}
\end{equation}
The shell formation time is always larger than the transition  time $t\rs{tr}$, provided
by the fact that $\eta>0$.
The ratio
\begin{equation}
 {t_1\over t\rs{tr}}=\eta^{1/(1+\eta)}
 \label{t1_ttr}
\end{equation}
is also always larger than unity.
Note that these relations do not depend on abundance and $\gamma$.
The ratios between all other times may be found from (\ref{t*-ttr}).

The consequence of times is $t\rs{dyn}<t\rs{tr}<t_1<t\rs{sf}$  (Fig.~\ref{fig-b}) in
nonuniform medium with increasing density. The time $t_1$ may be smaller than $t\rs{tr}$
and $t\rs{dyn}$ for the decreasing density medium. The sag time $t\rs{sag}<t\rs{tr}$ for
$\omega>-6$ only.

Fig.~\ref{fig-b} shows the two ratios (\ref{tsf_ttr}) and (\ref{t1_ttr}) as a functions
of $\omega$ for two values of $\beta$. Namely, the ratios $t_1/t\rs{tr}\approx 1.3$ and
$t\rs{sf}/t\rs{tr}\approx1.6\div 1.8$ are almost the same for shock in the medium with
increasing density ($\omega\leq 0$). Therefore, in case of a uniform medium and a medium
with  increasing density, {\em there is a need of introduction of transition subphase
with duration more than a half of SNR age at the beginning of this subphase}, $t\rs{tr}$.
The transition time $t\rs{tr}$ and therefore the transition subphase
$t\rs{sf}-t\rs{tr}\propto t\rs{tr}$ are less for higher density and lower initial energy:
$t\rs{tr}\propto E_{51}^{(2+2\beta+\omega)/\delta}A^{-(7+2\beta)/\delta}$ where
$\delta=11+6\beta-\omega(5+2\beta)$. Such dependence on density is also visible in
numerical calculations (Fig.~8 in \cite{Blondin}).

\begin{figure}
\centering
\includegraphics{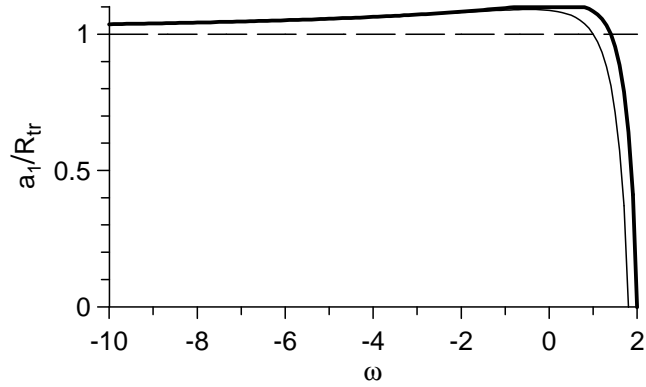}
\caption{The ratio $a_1/R\rs{tr}$ and $a\rs{dyn}/R\rs{tr}$  for $\beta=1$ (thick lines)
and $\beta=1/2$ (thin lines).} \label{fig-a1}
\end{figure}

\subsubsection{Medium with decreasing density}

It seems that the formulae (\ref{tsf_ttr}) and (\ref{t1_ttr})  suggest for the case of
decreasing density that the PDS radiative stage can even begin right after the end of
adiabatic stage: $t\rs{sf}/t\rs{tr}\to 0$ with $\omega\to 3(1+\beta)/(2+\beta)$. Another
result, already stated in \cite{Franko-et-al-94}, also follows: there will be no
radiative shell formation for $\omega\geq 3(1+\beta)/(2+\beta)$. In order to understand
the reasons of such behavior let us consider more details.

What is the coordinate $a_1$ of the element which cools first?  This element was shocked
at $t_1=\eta^{1/(1+\eta)}t\rs{tr}$. The Sedov radius at this time is $R(t_1)=a_1=
\eta^{2/\left((5-\omega)(1+\eta)\right)}R\rs{tr}$, thus the coordinate $a_1>R\rs{tr}$  if
$\omega<1.4$ ($\beta=1$) as it is shown on Fig.~\ref{fig-a1}. The ratio $a_1/R\rs{tr}$ is
close to unity and is almost the same for such $\omega$, i.e. the fluid we are interested
in will be shocked soon after $t\rs{tr}$. However, if $\omega>1.4$ then $a_1\to 0$
quickly with increasing of $\omega$ from 1.4 to 2, i.e. the element which cools first is
already inside the shock and may be in a very deep interior. The situation looks like
that there could not be any ``radiative shell'' in a common sense.

It is clear that the trend $t\rs{sf}/t\rs{tr}\to 0$ does not mean  that radiative
processes in the shock develop quickly for $\omega>1.4$. The transition and the shell
formation times correspond to different processes: $t\rs{tr}$ comes from comparison of
the initial thermal energy density of the shocked fluid with radiative losses though
$t\rs{sf}=t\rs{cool}$ is a time when the first cooled element appears. The two mentioned
processes have place in vicinity of the shock if ambient medium is uniform or with
increasing density. Numerical results suggest that they may be used for approximate
estimates of the limits of the transition subphase in such media. However these two
process are separated in space for media with decreasing density. It could be, that one
(or both) of the times $t\rs{tr}$ and $t\rs{sf}$ may not be suitable to mark stages of
SNR in medium with decreasing density.

The cooling of shock moving in the medium with decreasing density  differs from a
commonly accepted scenario of the ``thin dense shell'' formation and should be studied in
more details in the future.

\section{Conclusions}

The common approximate scenario of SNR evolution consists of the  free expansion stage,
the adiabatic phase and the PDS radiative era. It is shown that it is necessary to
consider also additional subphase between the adiabatic and the  radiative stages because
this subphase lasts more than half of SNR age it has at the end of adiabatic stage.

The analytical estimations on the ratios between the reference  times which characterize
the transition of adiabatic SNR into the PDS radiative stage -- $t\rs{tr}$, $t\rs{sf}$
and $t_1$ -- does not depend on the initial parameters of SNR and IMS (energy of
explosion, number density in the place of explosion, $\gamma$ etc.) except of the density
gradient (i.e. $\omega$) and assumed $\beta$ which causes rather small effect. This
result is also visible in the numerical calculations for case of the uniform medium
(Fig.~8 in \cite{Blondin}): except of the oscillations (which is indeed different for
different $n\rs{o}$) the durations of the transition subphase in terms of the transition
time are almost the same for different values of ISM density.

The ratio $t\rs{sf}/t\rs{tr}\approx 1.6$ for shock running in media with  constant or
increasing densities. The transition time however depends on the energy of explosion, the
density of the medium and the density gradient: $t\rs{tr}\propto
E\rs{SN}^{a(\omega)}A^{-b(\omega)}$ with $a>0$ and $b>0$ for shock in a medium with
$\rho\rs{o}\propto R^{-\omega}$. This means that the transition subphase is longer for
higher explosion energy and smaller density. The dependence of $t\rs{tr}$ on this
parameters are stronger for higher  $\omega$ because the functions $a(\omega)$ and
$b(\omega)$ increase with $\omega$.

The hydrodynamical properties of the shock in media with $\omega>0$  seem to cause a
trend to absence of the radiative phase in a common sense. The cooling of such shocks
differs from a commonly accepted scenario of the ``thin dense radiative shell'' formation
and should be studied in more details because it is important for models of nonspherical
SNRs which could be only partially radiative.

{\bf Acknowledgements.} {\em I am grateful to B.Hnatyk for valuable discussions.}

\section*{Appendix 1. \\ Approximation of the temperature evolution in a
 given fluid element downstream close to the strong adiabatic shock}

In order to simplify the estimation of $t\rs{sag}$ and $t\rs{dyn}$,  let us approximate
the distribution $\bar T(\bar a)=T(a,t)/T\rs{s}(t)$ downstream close to the strong
adiabatic shock; here $a$ is Lagrangian coordinate, $\bar T=T/T\rs{s}$ and $\bar a=a/R$.
Note, that hereafter in this Appendix we use the normalized parameters, i.e. divided on
their values on the shock front; thus we skip the overlines in the notations. We are
interested in the approximation in the form
\begin{equation}
 {T(a)}\approx a^{-\kappa(\gamma,\omega)}.
 \label{ap-base}
\end{equation}
The value of $\kappa$ is given by
\begin{equation}
 \kappa= \left(-{\partial \ln T(a)\over \partial \ln a}\right)_{a=1}
 \label{kappa-def}
\end{equation}
where $T(a)$ is the profile from Sedov \cite{Sedov-59} solutions.  The equation of the
mass conservation and the  equation of the adiabaticity applied for the case of the shock
motion in the medium with the power-law density distribution give the distribution of
temperature ${T}({a})={P}({a})/{\rho}({a})$ \cite{Petr-00}
\begin{equation}
 T(a)=\left({\gamma-1\over\gamma+1}\right)^{\gamma-1}
 a^{2\gamma-5+\omega} \left(r(a)^2r\rs{a}(a)\right)^{-\gamma+1}
 \label{Ta}
\end{equation}
where $r$ is Eulier coordinate and $r\rs{a}=\partial r/\partial a$. Instead  of Sedov
profiles for $r(a)$ -- which is quite complex -- we use the approximation
\begin{equation}
 r(a)=a^{(\gamma-1)/\gamma}\exp\big(\alpha(a^{\beta}-1)\big)
 \label{ra-app}
\end{equation}
where $\alpha,\beta$ are constants; this approximation gives correct  values of $r$ and
its derivatives in respect to $a$ up to the second order on the shock \cite{Petr-00}.
Substitution (\ref{kappa-def}) with (\ref{Ta}), (\ref{ra-app}) and with expressions for
$\alpha,\beta$ from \cite{Petr-00} yields
\begin{equation}
 \kappa={2\big(8-(\gamma+\omega)(\gamma+1)\big)\over(\gamma+1)^2}.
\end{equation}
For $\gamma=5/3$, $\kappa=1-3\omega/4$.

The approximation (\ref{ap-base}) underestimate Sedov temperature.  The smaller $a$ the
larger difference. It is about $20\%$ at $a\approx 0.5$ (that corresponds to
$r\approx0.8$).

\section*{Appendix 2. \\ List of times}

\timesss{$t\rs{sag}$}{``sag'' time \cite{Cox-72}, radiative cooling begins to affect the temperature distribution downstream of the shock;}
\timesss{$t\rs{dyn}$}{``dynamics-affected'' time \cite{Cox-72}, the temperature of a fluid element shocked after this time decreases faster due to radiation than due to expansion;}
\timesss{$t\rs{tr}$}{``transition'' time \cite{Blondin}, estimation of the time when the deviations from Sedov solutions are prominent; Sedov solution may be approximately used till this time;}
\timesss{$\Delta t\rs{cool}$}{``cooling'' time \cite{Kahn-1976,Cox-Anderson-82,Cox-86}, a shocked fluid element cools during this time;}
\timesss{$t\rs{s}$}{``shock'' time \cite{Cox-Anderson-82,Cox-86}, moment when the shock encountered given fluid element;}
\timesss{$t_1$}{moment when the shock encountered the fluid element which cools first \cite{Cox-Anderson-82,Cox-86};}
\timesss{$t\rs{c}$}{sum of $t\rs{s}+\Delta t\rs{cool}$;}
\timesss{$t\rs{cool}$}{``SNR cooling'' time \cite{Cox-Anderson-82,Cox-86,Cox-et-al-99}, the minimum of $t\rs{c}$, i.e. the age of SNR when the first cooled element appears;}
\timesss{$t\rs{sf}$}{``shell-formation'' time \cite{Cox-Anderson-82,Cox-86,Cox-et-al-99}, approximately after this time the shock may be described by the radiative PDS model;}
\timesss{$t\rs{low}$}{moment during the adiabatic stage when the radiative losses of the decelerating shock wave reach their minimum value;}
\timesss{$t\rs{hi}$}{moment when the radiative losses of the decelerating shock wave reach their maximum value \cite{Blinn-et-al-82,Silich-Pasko-86};}
\timesss{$t\rs{i}$}{``intersection'' time \cite{Band-Petr-04}, moment when two functions -- adiabatic $R(t)$ and radiative $R\rs{sh}(t)$ intersect;}
\timesss{$t\rs{max}$}{moment during the radiative stage when the function $m(\tau)$ reaches its maximum \cite{Band-Petr-04};}
\timesss{$\tilde t$}{timescale;}
\timesss{$\tau$}{dimensionless time, $\tau=t/\tilde t$.}


\end{document}